\documentclass{article}     
\usepackage[british]{babel}
\usepackage[utf8]{inputenc}
\usepackage{fmtcount}
\usepackage{amsthm}
\usepackage{verbatim}
\usepackage{hevea}

\newtheorem{defn}{Definition}
\newtheorem{alg}{Algorithm}
\newtheorem{example}{Example}
\mathchardef\mhyphen="2D
\loadcssfile{../../../css/bootstrap.css}
\begin{document}
\begin{htmlonly}
  \begin{rawhtml}
    <div class="navbar navbar-inverse navbar-fixed-top">
      <div class="navbar-inner">
	<div class="container">
          <ul class="nav">
            <li class="active"><a href="../../..">Home</a></li>
          </ul>
	</div>
      </div>
    </div>
    <div class="container">
  \end{rawhtml}
\end{htmlonly}
\title{On the Provenance of Linked Data Statistics}
\date{October 31\fmtord{st} 2010}
\author{William Waites\\
  \texttt{wwaites@tardis.ed.ac.uk}\\
  School of Informatics, University of Edinburgh\\
  Open Knowledge Foundation
}
\maketitle 
\begin{htmlonly}
  \begin{center}
  \begin{rawhtml}
    <p><em>alternate formats: 
        <a href="ldstat.txt">txt</a>,
        <a href="ldstat.pdf">pdf</a>
    </em></p>
  \end{rawhtml}
  \end{center}
\end{htmlonly}
\begin{abstract}
  As the amount of linked data published on the web grows,
  attempts are being made to describe and measure it. However
  even basic statistics about a graph, such as its size, 
  are difficult to express in a uniform and predictable
  way. In order to be able to sensibly interpret a statistic
  it is necessary to know how it was calculate. In this paper
  we survey the nature of the problem and outline a strategy
  for addressing it.
\end{abstract}

\section{Background and Motivation}

For the past several years datasets of Linked Open Data on
the web have been catalogued and made into a diagram 
\cite{LODC10} to illustrate their proliferation and 
interconnectedness. More recently some statistics about
these datasets have been calculated \cite{LODS10}. 
Amongst the published statistics are, for example the number
of triples in various graphs or unions of graphs across
a particular domain of interest. Some more sophisticated
statistics are also given in absolute terms, e.g. the
absolute number of links outgoing from a particular
dataset.

In conjunction with this work, a vocabulary \cite{VOID10}
has been developed for describing RDF datasets. This
vocabulary contains predicates for describing common
statistics, for example the number of triples or number
of distinct subjects, as well as some more generic
facilities for annotating a dataset description with
other types of statistical information.

Inasmuch as these statistics help to understand some
of the properties of these data at a coarse grained
level and get a rough idea of their dimensions they are
quite useful and indeed valuable contributions. However
as always we must ask what they mean. As, for example,
\texttt{void:triples} denotes the size of the dataset,
intuitively we might think that this gives some idea
of the amount of information contained in it. But,
most datasets carry a greater or lesser amount of
redundant information. It might be included to make
querying easier or to make extracts more easily
readable by a human. In some sense it could be 
argued that when counting triples that this redundant
information should be left out. It is also easily
demonstrated that an unlimited amount of redundant
triples can easily be added to any dataset without
really changing the information content. Clearly
the meaning of \texttt{void:triples} is somewhat of
a moving target.

The problem is exacerbated when more sophisticated
statistics are calculated. In the example of the 
outgoing links from one dataset to another, one might
want to normalise their count by dividing by the 
size of the dataset -- to arrive at a measure that 
might be called ``out-link density''. Perhaps such
a measure would tell us something about the character
of the dataset itself, independent of its size. But
because this measure is built on the basic notion 
of the size of the dataset we need to do some work
before arriving at a meaningful value for it.

\begin{quote}
  {\bf Notation Conventions:} Throughout the following,
  where RDF data is explicitly represented, the Notation 
  3 \cite{N3} syntax is used and declarations for common
  namespaces such as {\em rdf}, {\em rdfs}, {\em owl},
  {\em foaf}, {\em dct} are ommitted. In addition RDF
  terms and statements are represented in a fixed-with
  font.
\end{quote}

\section{Redundancy in Graphs}

A well known trivial example of adding redundancy to
graphs uses blank nodes. Blank nodes are to be read as
existential variables \cite{RDFS04}. 

\begin{example}{\bf Production Rules}
\end{example}

If we start with a graph containing one statement,
\smallskip
\texttt{
  \begin{center}
    bob a foaf:Person.
  \end{center}
}
\smallskip
we can then add a statement with a blank node,
\smallskip
\texttt{
  \begin{center}
    \_:b1 a foaf:Person.
  \end{center}
}
\smallskip
which simply says ``there exists someone that is a
\texttt{foaf:Person}''. This information was already 
contained in the original graph and really adds nothing
new. And we can add as many more blank nodes,
\texttt{\_:b2, \_:b3, ..., \_:bn}. What does this do to
a count of the size of the graph?

Such a production rule that generated these redundant
triples would called {\em unsafe} \cite{RIF10} but it is
quite possible to add a finite amount redundant
information with safe rules as well. The question of
how much redundant information has been added remains.

\begin{example}{\bf Graph Reduction}
\end{example}

As more realistic example, elaborated with the opposite
strategy of removing redundancy, consider the following
graph, and description logic fragment:
\smallskip
\texttt{
  \begin{center}
    \begin{tabular}{l|l}
      $s_1$ & bob a foaf:Person. \\
      $s_2$ & bob foaf:knows alice. \\
      $s_3$ & alice a foaf:Person. \\
      $s_4$ & alice foaf:knows bob. \\
      \hline
      $s_5$ & foaf:knows rdfs:domain foaf:Person.\\
      $s_6$ & foaf:knows rdfs:range foaf:Person.
    \end{tabular}
  \end{center}
}
\smallskip
  
If this is accompanied by the RDF semantic rules
\cite{RDFS04} (Section 4) concerning domains and 
ranges, namely,
\smallskip
\begin{center}
  \begin{tabular}{l|l}
    $r_1$ & \{\texttt{?s ?p ?o. ?p rdfs:domain ?A}\} \\
    & $\;\;\;\;$ \texttt{=>} \{\texttt{?s a ?A} \}. \\
    $r_2$ & \{\texttt{?s ?p ?o. ?p rdfs:range ?B}\} \\
    & $\;\;\;\;$ \texttt{=>} \{\texttt{?o a ?B} \}.
  \end{tabular}
\end{center}
\smallskip
we can immediately see that, under the given rules, statements
$s_1$ and $s_3$ are redundant as they can be derived from the
statements involving the predicate \texttt{foaf:knows} and
knowledge about its domain and range.

We need to make some distinction here between the graph under
consideration, ${ s_1, ..., s_4 }$ and the extra information
we have drawn upon, ${ s_5, s_6 }$. In general the latter will
come from a source external to the former, being referenced
as a vocabulary. It can also easily be seen that the description
logic fragment could just as well be expressed as a rule 
itself,
\smallskip
\begin{center}
  \begin{tabular}{l|l}
    $r_3$ & \{\texttt{?a foaf:knows ?b}\} \\
    & $\;\;\;\;$ \texttt{=>} \{\texttt{?a a foaf:Person.} \\
    & $\;\;\;\;$ \texttt{$\;\;\;\;\;$ ?b a foaf:Person.}\}
  \end{tabular}
\end{center}
\smallskip
  
We can further see that if we have $r_3$, we don't need $r_1$ 
or $r_2$.

So far we have shown that for,
\smallskip
\begin{center}
  \begin{tabular}{ll}
    $G$ & $= \{ s_1, ..., s_4 \} \cup \{ s_5, s_6 \}$ \\
    ${\mathcal R}$ & $= \{ r_1, r_2 \}$ \\
    $G^\prime$ & $= \{ s_2, s_4 \}$ \\
    ${\mathcal R}^\prime$ & $= \{ r_3 \}$
  \end{tabular}
\end{center}
\smallskip
the $G$ and $G^\prime$ under $\mathcal{R}$ and $\mathcal{R}^\prime$
respectively are in some sense equivalent, though it remains to state
explicitly what is meant by that. By inspection we can see that $G^\prime$
is half the size of $G$, in terms of number of statements, so we might say
that $50\%$ of $G$ was redundant. It is not immediately clear if
$\mathcal{R}$ and  ${\mathcal R}^\prime$ are the same or different
sizes.

\section{Rules and Redundancy}

The examples above rely on rules or what are often called
entailment regimes. Some common ones are defined in a 
placeholder vocabulary by the W3C \cite{SWER10}. Rules may
be applied in the usual way, as in the first example, to
produce statements and this is known as calculating the
closure of the graph.

\begin{defn}
  The {\em closure} of a graph, $G$ with respect to a set of
  rules, ${\mathcal R}$ is the set of all statements produced by
  applying ${\mathcal R}$ to $G$ until exhaustion, in other words
  adding all statements that it is possible to infer given the
  data and the rules. We will denote this operation as 
  $G^{{\mathcal R}+}$.
\end{defn}

Calculating $G^{{\mathcal R}+}$ is computationally expensive
but tractable with safe rules. Note that the closure with
respect to an empty ruleset is just the graph itself, or
$G = G^{\emptyset+}$.

\begin{defn}
  The {\em cardinality} of a graph, $G$ is simply the number of
  triples it contains and is denoted $|G|$.
\end{defn}

This is enough to provide a stable notion of the size of 
the graph by adding in a predictable proportion of redundancy
by specifying the entailment regime. Comparing the cardinality
of two graphs side-by side can thus be done in a meaningful
way.

\section{The Minimisation Problem}

What might be better than considering graphs which have had
redundancy added, however predictable it might be, is to
consider graphs with all possible redundancy eliminated. This
question was first considered in \cite{MEIER08} where Meier
proposed applying rules negatively.

\begin{alg}{{\bf Meier's Algorithm}\footnote{
  Meier didn't explicitly state the use of backwards 
  chaining but simply said to check if a given triple
  was possible to infer from the remaining triples
  under the given rules. The formulation given here
  comes from an actual implementation by the author
  of the present paper \cite{GM10}.
}}
  \smallskip
\begin{verbatim}
def reduce(graph, rules):
    for triple in graph:
        graph.remove(triple)
        if not backchain(graph, rules,
                         goal=triple):
            graph.add(triple)
    return graph
\end{verbatim}
  \smallskip
\end{alg}

The problem was further considered by Polleres et al. in
\cite{MINI10} and given the name {\em MINI-RDF}. The 
formulation builds on Meier's work and asks how 
difficult it is to find an irreducible graph given a
set of rules and a set of constraints\footnote{
  They also considered the problem of rule reduction
  in addition to graph reduction, the general case
  of the simplifications to the rules given in
  the second example above.
}. Constraints are simply rules that specify that a
certain amount of redundancy must be left in the graph
when a reduction has been completed. It turns out that
the problem is not tractable in general.

Whilst Meier's algorithm can be completed in polynomial
time, it will only find an irreducible graph that 
entails the same closure of the original. It was 
proven in \cite{MINI10} that finding the {\em smallest}
irreducible graph is in general also intractable.

\begin{defn}
  A {\em minimisation} of a graph, $G$ with respect to
  a set of rules, $\mathcal{R}$ is the smallest possible
  graph that has the same closure with respect to 
  $\mathcal {R}$ as $G$ and is written,
  $G^{\mathcal{R}-}$. $G^{\mathcal{R}-}$ is a solution
  to $\mathrm{MINI \mhyphen RDF}(G, \mathcal{R}, \emptyset)$.
\end{defn}

\begin{example}
\end{example}
It can be easily seen that $G^{R-}$ is not, in general,
unique. For example, consider the following graph,
$G = \{ s_1, s_2 \}$ and rule, ${\mathcal R} = \{ r_1 \}$,
\smallskip
\begin{center}
  \begin{tabular}{l|l}
    $s_1$ & \texttt{a links\_to b.} \\
    $s_2$ & \texttt{b linked\_from a.} \\
    \hline
    $r_1$ & \{\texttt{?x links\_to ?y}\} \\
    & $\;\;\;\;$ \texttt{<=>} \{\texttt{?y linked\_from ?x}\}
  \end{tabular}
\end{center}
\smallskip
either of $s_1$ or $s_2$ could be deleted from $G$ to obtain
$G^{\mathcal{R}-}$. 

Since finding $G^{\mathcal{R}}$ is in general intractable, a
potentially fruitful avenue of future research is to consider
the circumstances under which it can be solved in polynomial
time. Obviously the trivial case, $G^{\emptyset -}$ is 
solvable. For some kinds of rules, such as those in the
second example above, the minimisation will be unique. For
a broader set of rules, such as those with loops as in the
third example above, $|G^{\mathcal{R}-}|$ will be unique.
This last, the set of cardinality-preserving rules for
which calculating the minimisation of a graph is tractable,
is the broadest set of interest for the present purposes
-- rules in this class are practical to apply to remove as
much redundancy as possible from the graph.

\section{Redundancy Revisited}

We are now in a position to make some formal definitions of
what we mean by redundancy in graphs.

\begin{defn}
  The {\em redundancy} contained in a graph, $G$ with respect
  to a set of rules, ${\mathcal R}$ is given by
  $1 - {|G^{{\mathcal R}-}}|/{|G|}$.
\end{defn}

The foregoing considerations give rise to four fundamental
statistics about a graph, given a set of rules,

\begin{itemize}
  \item The cardinality of the graph as published.
  \item The cardinality of the closure of the graph under a given
    set of rules.
  \item The cardinality of a minimal graph under a given set of
    rules.
  \item The redundancy of the graph under a given set of rules.
\end{itemize}

From these it is possible to build up more elaborate statistics
in a predictable way. To take the ``out-link density'' example
from the introduction, this might be expressed as,
$$
D^{+}_{out}(G) = \frac{|G^{{\mathcal R}+}_{out}|}{|G^{{\mathcal R}+}|}
$$
or alternatively,
$$
D^{-}_{out}(G) = \frac{|G^{{\mathcal R}-}_{out}|}{|G^{{\mathcal R}-}|}
$$
Where the numerator has been constructed by selecting triples 
whose objects are resources in a different graph from the minimised
graph. If the rules are as strong as possible, such a statistic
might tell us something charactistic of the graph, if it is closer
to $0$ the graph contains mostly internal information as might be
the case with large datasets such as DBpedia or OpenCyc. A
\texttt{void:Linkset} on the other hand might have a characteristic
out-link density closer to $1$ as most of its statements express
the relationships between other datasets.

\section{Vocabulary Considerations}

The fact that the three of the fundamental statistics depend on the
rules used means that in order to express them unambiguously we need
also to mention the rules or entailment regimes. This is not a large
burden but does mean that we need a vocabulary for it. Such a
vocabulary would need to have predicates for including both Horn rules
and description logics. As support for the Rule Interchange Format
\cite{RIFBLD10} becomes more common it will be necessary to 
include rules expressed in this language as well.

The placeholder vocabulary for entailment regimes \cite{SWER10}
is a good starting point. The URIs defined there
are useful as recognisable unique identifiers but as yet have no
formal descriptions beyond pointers to the human readable 
documentation -- there is no automated way to discover which
rules each regime implies.

The \cite{SPIN09} vocabulary could be adapted for this but it
relies heavily on modelling SPARQL CONSTRUCT queries. While it 
has been shown \cite{SR07} that there is a mapping from these
types of queries to FOPL, most rulesets aren't written this 
way and it doesn't make so much sense to map rules from their
native representation to this vocabulary simply in order to
indicate their use.

We therefore propose a lightweight vocabulary \cite{GN10} for
Graph Normalisation, \texttt{gn}, as exemplified below\footnote{
  Recent work on the voiD vocabulary tends to deprecate the
  \texttt{void:statItem} predicate and the use of the SCOVO 
  vocabulary for expressing statistics. In our view this
  mechanism should be retained or replaced with something
  similar to support the expression of statistical provenance.
},
\smallskip
\begin{example}{\bf Graph Normalisation Vocabulary}

\begin{verbatim}
eg:dataset a void:Dataset ;
  void:statItem [ 
    scovo:dimension eg:redundancy ;
      rdf:value 0.3 ;
      gn:normalisation [
        a gn:MiniRDF ;
          gn:rules [
            a gn:RuleSet ;
            gn:n3 <.../rdfs-rules.n3> ;
            gn:dlogic owl:, foaf: ;
            gn:rif <.../rif-rules.rif>
          ]
      ]
  ].
\end{verbatim}
\end{example}

Additionally, \texttt{gn:constraints} is defined for completeness
to support specification of the {\em MINI-RDF} problem though in
practice this would probably never be used.

In this way, to check or recreate this statistic one
might procede from such a description as follows,

\begin{example}{\bf Redundancy Calculation}
\end{example}

\begin{enumerate}
  \item Fetch the dataset in question, $G$.
  \item Fetch Horn (N3, RIF) rules.
  \item Fetch description logics, making sure to follow
    \texttt{owl:imports}, collectively $D$.
  \item Transform the description logics into their equivalent
    Horn rules.
  \item Construct $\mathcal{R}$ as the set all the rules fetched.
  \item Run Meier's algorithm using $G \cup D$ as the graph for
    the backchaining step.
  \item Compute $1 - |G^{\mathcal{R}-}|/|G|$.
\end{enumerate}

This example of course assumes that the all rules used are
tractable for $G^{\mathcal{R}-}$ and cardinality preserving.

It should be noted that the current practice with respect to
\texttt{void:triples} is simply the above with an empty
ruleset. In this way some amount of backwards compatibility
with current practice is maintained.

\section{Optimising with Graph Diffs}

If finding a $G^{{\mathcal R}-}$ is tractable, it is still a
computationally expensive operation. If provenance information is
kept for datasets such that given the previous version and the
provenance metadata it is possible to reconstruct the current
version there is a significant optimisation to be had, especially
for large datasets that experience incremental change.

Starting with a minimisation of the previous version,
$G^{{\mathcal R}-}_i$, and a pair of graphs $I$ and $D$
representing triples to be inserted or deleted\footnote{
  Assuming blank nodes are handled via some sort of 
  skolemisation mechanism.
} such that $G_{i+1} = G_i - D + I$, we can construct
$G^{{\mathcal R}-}_{i+1}$ by first calculating,
$G^{{\mathcal R}-}_i - D$ since if they aren't in $G_{i+1}$
they won't be in its minimisation. This intermediate 
graph is a possibly non-minimal subgraph of $G_{i+1}$.
Because the order of the triples in \texttt{minimise()}
doesn't matter we can now run it only testing the triples
in $I$.

\section{Conclusion}

We have reviewed the truism that in order to be able
to sensibly interpret statistics one must know how they
are calculated. Looking at how this applies to descriptions
of RDF graphs we have seen that apparently simple statistics
can be calculated in a number of ways. Thus the importance
of provenance {\em of the statistics} has been highlighted.
A proposal for how this provenance information might be
expressed was put forward and some interesting areas for
further theoretical research were noted.

\bibliographystyle{plain}
\bibliography{ldstat}
\begin{htmlonly}
  \begin{rawhtml}
    </div>
  \end{rawhtml}
\end{htmlonly}
\end{document}